\documentclass[12pt,thmsa]{article}
\usepackage{sw20lart}

\input{tcilatex}
\begin{document}

\title{Analysis of a recent experimental test of Bell\'{}s inequalities violating
quantum predictions}
\author{Emilio Santos \\
Departamento de F\'{i}sica. Universidad de Cantabria. Santander. Spain}
\maketitle

\begin{abstract}
A recent experiment by Brida et al. (quant-ph/07050439) is analyzed with the
conclusion that it shows a significant violation of standard quantum
predictions. A simple local hidden variables model is studied which is
compatible with the empirical results and fits fairly well the deviation
from the quantum predictions.

\end{abstract}

\section{\protect\smallskip Introduction}

A recent experiment by Brida, Genovese and Piacentini\cite{Brida} has shown
a violation of an inequality which I derived\cite{Santos} for a restricted,
but sensible, family of local hidden variables (LHV) theories. The empirical
results, however, also violate the quantum predictions. Thus it is worth
studying more carefully the implications of the experiment, which is the
purpose of this paper.

As is well known, many experiments have been performed in the attempt to
discriminate between quantum mechanics and LHV theories via tests of
Bell\'{}s inequalities. The experiments have agreed with quantum mechanics
in general, but none of them has provided a conclusive, loophole-free,
refutation of the whole family of LHV theories. This is due to the fact that
genuine Bell inequalities, derived from locality and realism alone, are
extremely difficult to test\cite{S2006}. In fact the perfomed experiments
have tested Bell type inequalities derived from local realism plus some
additional assumptions. Thus the violation of the inequalities has refuted
restricted families of LHV theories, namely those fulfilling the auxiliary
assumptions. In the early experiments the additional assumption was ``no
enhancement''\cite{Clauser}, and the experiments provided a clear empirical
refutation of that family. Later on, for about 25 years beginning with the
Aspect experiments\cite{Aspect}, the additional assumption has been ``fair
sampling''. LHV theories with fair sampling have been clearly refuted by
many experiments. For instance the experiment by Brida et al.\cite{Brida}
reports a violation by 48 standard deviations (see their eq.(15)). However I
think that the fair sampling assumption excludes ``a priori''all sensible
LHV theories\cite{S2006} and therefore the refutation of those LHV theories
has not too much relevance. For this reason I have started the search for
Bell type inequalities derived from local realism plus some assumptions more
reasonable than fair sampling, thus being able to provide tests of some
restricted, but sensible, families of LHV theories.

I shall consider specifically experiments measuring polarization correlation
of optical photon pairs like the one performed by Brida et al.\cite{Brida}.
In the experiment a source produces photon pairs, each member of the pair
traveling along one of two possible paths, each path ending in an
analyzer-detector system (named Alice and Bob, respectively). If the
polarization planes of the analyzers are determined by the angles $\phi _{1%
\text{ }}$and $\phi _{2\text{ }},$ respectively, the results of the
experiment may be summarized in two single rates, $R_{1}(\phi _{1\text{ }})$
and $R_{2}(\phi _{2\text{ }})$ , and a coincidence rate $R_{12}(\phi _{1%
\text{ }},\phi _{2\text{ }}).$ The detection rates divided by the production
rate, $R_{0},$ not measurable in the experiment, are the detection
probabilities that is 
\begin{equation}
p_{j}(\phi _{j\text{ }})=\frac{R_{1}(\phi _{1\text{ }})}{R_{0}}%
,\;p_{12}(\phi _{1\text{ }},\phi _{2\text{ }})=\frac{R_{12}(\phi _{1\text{ }%
},\phi _{2\text{ }})}{R_{0}},  \label{02}
\end{equation}
which are the quantities to be calculated from the theory, either a LHV
model or quantum mechanics. Following Bell, a LHV model consists of three
functions, $\rho (\lambda ),P_{1}(\lambda ,\phi _{1}),P_{2}(\lambda ,\phi
_{2}),$ where $\lambda $ stands for one or several hidden variables, such
that the detection probabilities could be obtained by means of the integrals 
\begin{equation}
p_{j}(\phi _{j\text{ }})=\int \rho (\lambda )P_{j}(\lambda ,\phi
_{j})d\lambda ,\;p_{12}(\phi _{1\text{ }},\phi _{2\text{ }})=\int \rho
(\lambda )P_{1}(\lambda ,\phi _{1})P_{2}(\lambda ,\phi _{2})d\lambda .\;
\label{03}
\end{equation}
The essential requirements of realism and locality imply that the said
functions fulfil the conditions 
\begin{equation}
\rho (\lambda )\geq 0,\int \rho (\lambda )d\lambda =1,\;0\leq P_{j}(\lambda
,\phi _{j})\leq 1.  \label{04}
\end{equation}
The experiment is compatible with local realism if there exists a LHV model
reproducing the results of the experiment, that is if one may find three
functions $\rho ,P_{1},P_{2}$ and a rate, $R_{0},$ such that the results, $%
R_{1}$, $R_{2}$ and $R_{12}$ are reproduced by eqs.$\left( \ref{02}\right) $
and $\left( \ref{03}\right) $. In particular, for the proof of compatibility
it is not necessary to make any analysis of the source or the
analyzer-detectors systems, which may be taken as ``black boxes''. Also we
should not make any assumptions about the signals produced in the source,
the word ``photon'' being here just a short for ``signal of whatever nature
produced in the source and able to propagate, until its arrival to Alice or
Bob, with velocity not higher than that of light''. For later convenience I
shall label $LHV0$ the whole family of local hidden variables theories so
defined (i. e. by eqs.$\left( \ref{02}\right) $ to $\left( \ref{04}\right)
.) $

The restricted family of theories which I have proposed elsewhere\cite
{Santos} reduces $\lambda $ to a set of just two angular hidden variables,
that is $\lambda \equiv \left\{ \chi _{1},\chi _{2}\right\} ,$ where $\chi
_{1}$ ($\chi _{2})$ is a polarization angle of the first (second) photon of
a pair, thus $\chi _{j}$ and $\chi _{j}+\pi $ representing the same
polarization. In addition I assume a specific dependence of $\rho (\chi
_{1},\chi _{2})$ and $p_{j}(\chi _{1},\phi _{j\text{ }})$ so that Bell\'{}s
eqs.$\left( \ref{03}\right) $ and $\left( \ref{04}\right) $ become

\begin{eqnarray}
p_{12}(\phi ) &=&\int \rho (\chi _{1}-\chi _{2})P(\chi _{1}-\phi _{1})P(\chi
_{2}-\phi _{2})d\chi _{1}d\chi _{2},  \label{1} \\
p_{j} &=&\int \rho (\chi _{1}-\chi _{2})P(\chi _{j}-\phi _{j})d\chi
_{1}d\chi _{2},\;j=1,2,  \label{1a}
\end{eqnarray}
with the conditions 
\begin{equation}
\rho (x)=\rho (-x)\geq 0,\int \rho (x)dx=1/\pi ,\;0\leq P(x)=P(-x)\leq 1.
\label{2}
\end{equation}
(The normalization of $\rho $ fulfils eq.$\left( \ref{04}\right) $ if we
integrate over both hidden variables, $\chi _{1\text{ }}$and $\chi _{2}.)$ I
shall label $LHV1$ the family defined by eqs.$\left( \ref{1}\right) ,$ $%
\left( \ref{1a}\right) $ and $\left( \ref{2}\right) .$

The main inequality derived for the family $LHV1$ is\cite{Santos}

\begin{eqnarray}
\Delta _{\exp } &\equiv &\left\{ \frac{1}{n}\sum_{k=1}^{n}\left[ \frac{%
R_{12}(\phi _{k})}{\left\langle R_{12}\right\rangle }-1-V\cos 2\phi
_{k}\right] ^{2}\right\} ^{1/2}\geq D(\eta )  \label{3} \\
D(\eta ) &\simeq &\frac{8\sqrt{2}}{3\pi }\sqrt{\frac{2}{3\eta }-\frac{1}{2}-%
\frac{\sin ^{4}\left( \pi \eta /2\right) }{\left( \pi \eta /2\right) ^{4}}}%
\varepsilon ^{3},\varepsilon \simeq \frac{1}{\sqrt{2}}\left( V-\frac{\sin
^{2}\left( \pi \eta /2\right) }{\left( \pi \eta /2\right) ^{2}}\right)
_{+}^{1/2}  \label{3b}
\end{eqnarray}
where $\phi _{k}=\pi k/n,$ $k=1,2...n,$ stands now for the difference
between the polarization angles $\phi _{1}$ and $\phi _{2}$, of Alice and
Bob respectively, and $\left( .\right) _{+}$ means putting zero if the
quantity inside the bracket is negative. The quanity $\eta $ enters in the
model as the ratio between twice the coincidence detection rate, $%
R_{12}(\phi ),$ averaged over angles and the single rate $R_{1}\simeq R_{2}$
or their mean if $R_{1}\neq R_{2}$ (assuming that the single rates do not
depend on the angles $\phi _{j},$ as is usual) and $V$ should be obtained
from the best cosinus fit to the empirical coincidence detection rates, that
is 
\begin{equation}
\left\langle R_{12}\right\rangle =\frac{1}{n}\sum_{k=1}^{n}R_{12}(\phi
_{k}),\;\eta =\frac{4\left\langle R_{12}\right\rangle }{R_{1}+R_{2}},\;V=2%
\frac{\sum_{k=1}^{n}R_{12}(\phi _{k})\cos 2\phi _{k}}{n\left\langle
R_{12}\right\rangle }.  \label{4a}
\end{equation}
The quantity $D(\eta )$ provides a lower bound for the deviation between the
best local model of the family $LHV1$ and quantum mechanics, but the
approximate expression eq.$\left( \ref{3b}\right) $ is valid only for low
detection efficiencies (see below).

According to eq.$\left( \ref{4a}\right) $ the quantity $\eta $ corresponds
to an overall detection efficiency, taking into account all kinds of losses
in lenses, polarizers, etc. But in typical experiments, including the one by
Brida et al.\cite{Brida}, the quantity $\eta $ so defined is rather small
with the consequence that the inequality $\left( \ref{3}\right) $ is very
well fulfilled, thus making the tests of the $LHV1$ familly vs.quantum
mechanics almost impossible. In consequence I have proposed to restrict the
family $LHV1$ by including a ``partial fair sampling'' assumption which
applies to lenses, polarizers, etc.\textit{\ but not to the detectors }(see
my paper\cite{Santos}.) This means that the quantity $\eta $ to be used in
the inequality $\left( \ref{3}\right) $ should be the one given by eq.$%
\left( \ref{4a}\right) $ divided by the product $f_{1}f_{2}.....f_{s}$,
where 1, 2, ...s correspond to the different devices inserted bewteen the
source and the detectors, like lenses, filters, polarizers or even the
medium which transmits the photons, and $f_{l}$ is the fraction of photons
that are \textit{not} absorbed in the corresponding device. In practice this
is more or less equivalent to using for $\eta $ the quantum efficiency of
the detectors themselves, to be measured in auxiliary experiments.

\section{Tests of local hidden variables theories vs. quantum mechanics}

Now I shall study the specific experiment by Brida et al.\cite{Brida}. The
results of the experiment are summarized in the following table (not
published in the report of the experiment\cite{Brida}; I acknowledge the
authors for providing me with this valuable information)

\bigskip

\textbf{Table 1. Coincidence rates vs. angle amongst polarizers\bigskip }

${}$\noindent $\! 
\begin{array}{lllllllll}
\phi \text{ }(\deg ) & 0 & 22.5 & 45 & 67.5 & 90 & 112.5 & 135 & 157.5 \\ 
R_{12}(\phi ) & 9906.2 & 8439.6 & 4936.6 & 1454.1 & 108.0 & 1481.3 & 4983.5
& 8499.2 \\ 
\Delta R_{12} & 21.0 & 18.6 & 13.6 & 9.0 & 8.2 & 11.9 & 14.1 & 19.0
\end{array}
$

\smallskip \bigskip

Quantum mechanics predicts a cosinus curve of the form

\begin{equation}
\frac{R_{12}(\phi _{j})}{\left\langle R_{12}\right\rangle }=1+V\cos (2\phi
_{j}+\psi ),  \label{11}
\end{equation}
where the phase $\psi $ is included in order to take account of any possible
error in the measurement of the angle between polarizers. The best fit to
the results of Table 1 gives 
\begin{equation}
V=0.9897,\psi =0.31(\deg ).  \label{12}
\end{equation}
The fit is rather bad, predicting in particular 
\begin{equation}
R_{12}(90{{}^{o}})=\left\langle R_{12}\right\rangle \left[ 1+V\cos (180{%
{}^{o}}+\psi )\right] =51.3,  \label{12a}
\end{equation}
which has a significant deviation from the value given in Table 1. \textit{%
This shows that the quantum prediction eq.}$\left( \ref{11}\right) $\textit{%
\ is violated}, which may be also seen from the value 
\begin{eqnarray}
\frac{V_{B}}{V_{A}} &=&1.0205\pm 0.0048,  \label{12b} \\
V_{A} &\equiv &\frac{R_{12}(0{{}^{o}})-R_{12}(90{{}^{o}})}{R_{12}(0{{}^{o}}%
)+R_{12}(90{{}^{o}})},V_{B}\equiv \sqrt{2}\frac{R_{12}(22.5{{}^{o}}%
)-R_{12}(67.5{{}^{o}})}{R_{12}(22.5{{}^{o}})+R_{12}(67.5{{}^{o}})}, 
\nonumber
\end{eqnarray}
reported by Brida et al.\cite{Brida}. It is clearly incompatible with the
standard quantum prediction $V_{A}=V_{B}$ $.$

If we ignore the value $R_{12}(90{{}^{o}})$ of Table 1, a very good fit is
obtained to eq.$\left( \ref{11}\right) $ with 
\begin{equation}
V=0.9966,\psi =0.31(\deg ).  \label{13}
\end{equation}
Indeed this fit reproduces all values of Table 1 well within statistical
errors, except for 90${{}^{o}}$ where it predicts $R_{12}(90{{}^{o}})=17.0$,
to be compared with the value $108.0$ of Table 1. We see that the violation
of quantum predictions is due to the too high value of the empirical
coincidence counting rate at $\phi =90{{}^{o}}.$

In order to test the family of local models defined by eqs.$\left( \ref{1}%
\right) $ to $\left( \ref{2}\right) ,$ we must check whether the inequality $%
\left( \ref{3}\right) $ holds true and for this purpose we shall choose the
appropriate value of the parameter $\eta .$ The experiment\cite{Brida}
belongs to a class where half of the photons produced in the source are
excluded by a post-selection procedure. Experiments of this kind have been
performed since long ago\cite{Mandel}. However there has been some
controversy about whether these experiments actually allow tests of
Bell\'{}s inequalities\cite{Caro}. Indeed by the nature of the source only
half the photons produced belong to pairs going to different detectors (that
is one to Alice and the other one to Bob) so that the effective overall
detection efficiency cannot be larger than 50\%, that is much lower than the
minimum required for the violation of a (genuine) Bell inequality. Actually
experiments of this type \textit{do allow} Bell tests, but only if
two-channel analyzers followed by two detectors are used by Alice and
similarly by Bob, so that all photon pairs may (in principle) be detected%
\cite{Santos2}. In the experiment by Brida et al.\cite{Brida} Alice and Bob
possess only one detector each, so that it can be interpreted by a local
model like the one defined by eqs.$\left( \ref{1}\right) $ and $\left( \ref
{1a}\right) $. Furthermore, we must use a parameter $\eta $ with a value%
\textit{\ }just\textit{\ half the quantum efficiency of the actual detectors}%
, which in this experiment is quoted to be 0.62. Indeed the ratio between
twice the average coincidence rate and the single rate would be half the
quantum efficiency at most, the maximum taking place if there were no losses
bewteen the source and the detectors. The family of local models for the
said experiment\cite{Brida} with $\eta =0.31$ will be labelled $LHV2$.

It may seem plausible to interpret the experiment by assuming that photons
are particles and that the effect of the non-polarizing beam splitter is to
divide the ensemble of photon pairs arriving at it (coming from the
non-linear crystal) into three subensemble consisting respectively of photon
pairs going: 1) both photons to Alice, 2) both to Bob, 3) one of them to
Alice and the other one to Bob. Within this (corpuscular) model of light it
is appropriate to ignore the single rates due to photons such that both
members of the pair go to Alice or both to Bob, which are precisely half of
the photon pairs produced in the source. Thus we may consider LHV models
involving only the photon pairs of the third subensemble. If we add the
``partial fair sampling'' assumption, we are led to use an efficiency $\eta
=0.62$ in the inequality $\left( \ref{3}\right) .$ This defines a family of
models more restricted than $LHV2$ which I shall label $LHV3$. Obviously we
might consider also families intermediate between $LHV2$ and $LHV3$ or
between $LHV1$ and $LHV2,$ each of them fulfilling inequality $\left( \ref{3}%
\right) $ with intermediate values of $\eta ,$ but I shall not discuss this
possibility here.

A particular case of the family $LHV3$ is obtained using a density $\rho
(\chi _{1}-\chi _{2})$ of the form 
\begin{equation}
\rho (x)=\frac{1}{\pi ^{2}}\left[ 1+\left( 1+\varepsilon \right) \cos \left(
2x\right) +\varepsilon \cos \left( 4x\right) \right] ,\varepsilon \in \left[
0,\frac{1}{3}\right] ,  \label{5}
\end{equation}
which was studied elsewhere\cite{Santos1} and I shall label $LHV4.$ Thus I
have defined a hierarchy of families of local models 
\begin{equation}
LHV0\supset LHV1\supset LHV2\supset LHV3\supset LHV4.  \label{6}
\end{equation}

The family $LHV1$ cannot be tested without the knowledge of the single
detection rates, but we may safely claim that it is not refuted by the
experiment of Brida et al.. Indeed the value of $\eta $ derived from eq.$%
\left( \ref{4a}\right) $ would be very small. In contrast the family $LHV4$
has been clearly refuted. In fact an inequality derived from eq.$\left( \ref
{5}\right) $\cite{Santos1} is violated by more than 11 standard deviations%
\cite{Brida}$.$ However the question whether the families $LHV2$ and $LHV3$
have been refuted requires a more careful analysis, which is made in the
following.

Using a detection efficiency $\eta =0.62$ for the empirical test of eq.$%
\left( \ref{3}\right) $ the authors\cite{Brida} report a violation, by $3.3$ 
$\sigma ,$ of the inequality $\left( \ref{3}\right) .$ However they used for 
$D(\eta )$ the expression eq.$\left( \ref{3b}\right) $ which is valid only
for relatively low efficiency. Indeed in my article\cite{Santos} it is
stated that eq.$\left( \ref{3b}\right) $ (eq.(40) of my paper) is obtained
to order $\varepsilon ^{3}$ in the parameter $\varepsilon $ (defined in my
eq.(35).) In the experiment of Brida et al. $\varepsilon \simeq 0.43$ is not
small and, consequently, the exact eqs.(31), (34) and (38) of the paper\cite
{Santos} should be used, rather than the approximation to lowest order in $%
\varepsilon .$ (I apologyze for not having made this point more clear in my
article). Using eqs.(31) and (34) of my paper\cite{Santos} we get the
following equation for $\varepsilon $%
\begin{equation}
\frac{\pi -2\varepsilon +\sin (2\varepsilon )\cos (2\varepsilon )}{\cos
(2\varepsilon )\left[ \pi -2\varepsilon +\tan (2\varepsilon )\right] }=V%
\frac{\left( \pi \eta /2\right) ^{2}}{\sin ^{2}\left( \pi \eta /2\right) },
\label{14}
\end{equation}
whence, with the value eq.$\left( \ref{12}\right) $ for $V$ and $\eta =0.62,$
I obtain $\varepsilon =0.578.$ For such a high value the calculation to
lowest order in $\varepsilon $ is not valid, but an accurate lower bound of $%
D(\eta )$ is obtained by means of 
\begin{equation}
D(\eta )\geq \frac{\sqrt{2}\sin ^{3}\left( 2\varepsilon \right) }{3\left[
\left( \pi -2\varepsilon \right) \cos (2\varepsilon )+\sin (2\varepsilon
)\right] }\frac{\sin ^{2}\left( \pi \eta \right) }{\left( \pi \eta \right)
^{2}}=0.048.  \label{14a}
\end{equation}
(See eq.(39) of my paper\cite{Santos}). This gives rise to a violation of
the inequality $\left( \ref{3}\right) $ even stronger than the one reported%
\cite{Brida} so that the experiment clearly refutes the family $LHV3$.

As said above an effective efficiency $\eta =0.31$ should be used in the
test of the family $LHV2$ via inequality $\left( \ref{3}\right) .$ In this
case we may take $D\left( \eta \right) $ as given by eq.$\left( \ref{3b}%
\right) $ because the parameter $\varepsilon $ has the value $\varepsilon
=0.1820$, which is low enough for the approximations involved being valid
(the exact eq.$\left( \ref{14}\right) $ gives $\varepsilon =0.1825).$ Thus I
get from eq.$\left( \ref{3b}\right) $%
\[
D\left( \eta \right) =0.0065<\Delta _{\exp }=0.0074,
\]
that is the inequality $\left( \ref{3}\right) $of the family $LHV2$ is
fulfilled. (The lower bound eq.$\left( \ref{14a}\right) $ is now $D(\eta
)\geq 0.0052.)$

The model of the family $LHV2$ which is most close to quantum mechanics
predicts a deviation from the best cosinus fit (that is eq.$\left( \ref{11}%
\right) $ with $V=0.9897)$ of the form (see eq.(37) of my paper\cite{Santos}%
) 
\begin{eqnarray}
\delta \left( \phi \right)  &=&\alpha \left[ \beta \cos \left( 2\phi \right)
-1\right] +\gamma \left( \phi \right) ,  \label{15} \\
\alpha  &\equiv &\frac{8\varepsilon ^{3}}{3\pi },\beta \equiv 2\frac{\sin
^{2}\left( \pi \eta /2\right) }{\left( \pi \eta /2\right) ^{2}},\gamma
\left( \phi \right) =\frac{2\alpha }{\eta ^{2}}\left( \eta +\frac{2}{\pi }%
\left| \phi \right| -1\right) _{+},  \nonumber
\end{eqnarray}
where $\varepsilon $ was defined in eq.$\left( \ref{3b}\right) $, $\phi \in
\left[ -\frac{\pi }{2},\frac{\pi }{2}\right] $ and $\left( {}\right) _{+}$
means putting 0 if the quantity inside brackets is negative. With $\eta =0.31
$ the first two terms of eq.$\left( \ref{15}\right) $ predict an effective
increase in the parameter V leading to 
\[
V\rightarrow V_{eff}=\frac{V+\alpha \beta }{1-\alpha }\simeq V+\alpha \left(
1+\beta \right) =1.003
\]
which is somewhat larger than $V$ in the best cosinus fit to the data of
Table 1 when the value $R_{12}(90{{}^{o}})$ is excluded, see  eq.$\left( \ref
{13}\right) .$ The contribution of the last term of eq.$\left( \ref{15}%
\right) $ is either zero or negligible for all angles reported in Table 1
except $\phi =90{{}^{o}}$ (note that any angle $\phi >90{{}^{o}}$ in Table 1
should be replaced by $180{{}^{o}}-\phi $ if used in eq.$\left( \ref{15}%
\right) .)$ For 90${{}^{o}}$ we get $\gamma \left( 90{{}^{o}}\right) =0.0330$
$,$ $\delta \left( 90{{}^{o}}\right) =0.0184$ which, taking into account the
value $\left\langle R_{12}\right\rangle \simeq 4980$ obtained from Table 1,
gives a predicted increase $\Delta R_{12}(90{{}^{o}})=89.6$. If this is
added to the prediction of the fit eq.$\left( \ref{12}\right) ,$ we get a
LHV model prediction $R_{12}(90{{}^{o}})\simeq 140.9$, which is somewhat
larger than the empirical datum of Table 1. The results of our calculation
strongly suggest that a local model defined by eqs.$\left( \ref{1}\right) $
to $\left( \ref{2}\right) ,$ with a value of the parameter $\eta $ slightly
smaller than 0.31, may agree with the empirical results. But this point will
not be studied further in the present paper.

\section{Conclusions}

In summary the results of the experiment by Brida et al., shown in Table 1,
are not compatible with the standard predictions of quantum mechanics, eq.$%
\left( \ref{11}\right) .$ Nevertheless it may be that small corrections, not
included in the standard quantum calculations, might account for the
disagreement between theory and experiment. In contrast a sensible family of
local models\cite{Santos} predicts fairly well the empirical departure from
standard quantum mechanics, that is a substantial increase in the
coincidence counting rate at angles close to $\phi =90{{}^{o}}$.

I will finish stating that the experiment by Brida et al.\cite{Brida} is
remarkable in that it has achieved, for the first time to my knowledge, a
value of the parameter $V_{B}$ very close to unity ( the departure is only
1.5 per thousand) combined with a fairly high quantum efficiency of the
detectors. These properties are crucial for the discrimination between
standard quantum predictions and sensible families of local hidden variables
theories, like the one defined by eqs.$\left( \ref{1}\right) $ to $\left( 
\ref{2}\right) .$

\end{document}